\documentclass[pra,english,preprintnumbers,amsmath,amssymb,nofootinbib,superscriptaddress]{revtex4-1}
\usepackage{dirtytalk}
\usepackage{tikz}
\usepackage{subcaption}
\usepackage{tikz-feynman}
\tikzfeynmanset{compat=1.1.0}
\usepackage{amsmath}  
\usepackage{amsfonts} 
\usepackage{graphicx} 
\usetikzlibrary{shapes.misc}
\tikzset{cross/.style={cross out, draw=black, minimum size=2*(#1-\pgflinewidth), inner sep=0pt, outer sep=0pt},
	cross/.default={3.5pt}}
\newcommand{\pb}[1]{\psi_B(#1)}
\newcommand{\pk}[1]{\psi_k(#1)}
\begin{document}
	\title{ Spectral density, Levinson's theorem, and the extra term in the second virial coefficient for 1D delta-function potential
	}
	
	\author{H. E. Camblong}
	\affiliation{Department of Physics and Astronomy, University of San Francisco. San Francisco, CA 94117-1080 USA}
	
	\author{A. Chakraborty}
	\affiliation{Physics Department, University of Houston. Houston, Texas 77024-5005, USA}
	
	\author{W. S. Daza}
	\affiliation{Escuela de f\'isica, Universidad Pedag\'ogica y Tecnol\'ogica de Colombia (UPTC), Avenida Central del Norte, Tunja, Colombia.}
	
	\author{J. E. Drut}
	\affiliation{Department of Physics and Astronomy, University of North Carolina. Chapel Hill, North Carolina 27599-3255, USA}
	
	\author{C. L. Lin}
	\affiliation{Physics Department, University of Houston. Houston, Texas 77024-5005, USA}
	
	\author{C. R. Ord\'o\~nez}
	\affiliation{Physics Department, University of Houston. Houston, Texas 77024-5005, USA}
	\affiliation{ICAB, Universidad de Panam\'a, Panam\'a, Rep\'ublica de Panam\'a.}
	
	\begin{abstract}
	In contrast with the 3D result, the Beth-Uhlenbeck (BU) formula in 1D contains an extra $-1/2$ term. The origin of this $-1/2$ term is explained using a spectral density approach. To be explicit, a delta-function potential is used to show that the correction term arises from a pole of the density of states at zero energy. The spectral density method shows that this term is actually an artifact of the non-normalizability of the scattering states and an infrared cutoff regularization scheme has to be used to get the correct result in 1D. The formal derivation of the BU formula would miss this term since it ignores the effects of the boundary terms. While the result is shown for the delta-function potential, the method and result are valid for more general potentials. Additionally, the 1D Levinson's theorem can be extracted from the spectral density method using the asymptotic form of general potentials. The importance of the result lies in the fact that all these correction terms in 1D have a universal source: a pole at zero energy. Similar calculations using quantum field theoretical approaches (without explicit infrared cutoff regularization schemes) also show the same subtleties with the correction term originating from the zero energy scattering states (appendix \ref{appendixLeyronas}).
	\end{abstract}
	\date{\today}
	\maketitle
	

	\section{Introduction}
	The virial expansion was developed in the beginning of the 20th century \cite{onnes1901expression} to provide an estimate of the deviation of the behavior of real gases from the ideal gas equation. Since then, the second virial coefficient $b_2$ has been studied in great detail because it can be calculated analytically and gives the leading order correction for real gases. The virial coefficients reflect, order by order, the effects of interactions on the $n$-body problem: the $n$-th order virial coefficient is determined by solving the $n$-body problem. Therefore, one can obtain the shift in $b_2$ from the free particle case {\it analytically} by solving the 2-body Schr{\"o}dinger equation with an interaction potential. Beth and Uhlenbeck, in their famous paper~\cite{beth1937quantum}, showed that this shift $\delta b_2$ is related to the scattering phase shift in 3D. Since then the result has been generalized in lower dimensions as well \cite{portnoi1998levinson,kristensen2016second,cui2012quasi}. Recently, it was proven that $\delta b_2$ is the imprint of the quantum scale anomaly in dilute quantum gases in 1D with three-body and 2D with two-body local interactions \cite{daza2018virial,drut2018quantum} using a many-body path integral formalism \cite{ordpa}. In an effort to extend this work with a 1D local derivative-delta interaction, it was demonstrated that the derivative-delta potential shares a number of properties with the delta-function potential \cite{camblong2019quantum}. One of them is that a straightforward application of the Beth-Uhlenbeck (BU) formula misses a $-1/2$ term to give a \say{wrong} result in the limit $E_B \rightarrow 0$ ($E = -E_B$, $E_B>0$ is the bound state energy). It has been previously shown that in 1D, the correct BU formula has an additional $-1/2$ term which produces the correct result for the 1D delta-function in the limit $E_B\rightarrow 0$. The origin of this mysterious factor in 1D has been mentioned in several works \cite{amaya2005second,dodd1974cluster,sassoli1994levinson,barton1985levinson}. 
	
	\noindent The present article is an effort to make the formalism more understandable and to connect different contexts that share the same property with regards to the $-1/2$ term. In this note, we use the spectral-density approach to show in detail that this term is due to the non-normalizability of the scattering states and the contribution of the zero energy state of the system. We have used an infrared (large volume, length in 1D) cutoff regularization method to systematically control the divergences in the calculation, which gives us the desired $-1/2$ term. For simplicity, and for the sake of better understanding, we have used the results of scattering from a delta-function potential, but the method can be implemented for general potentials as well. In addition to providing the correction term in the BU formula, the spectral density approach can also be used to derive the 1D Levinson's theorem \cite{boya2007theorem}. We have also investigated similar subtleties using different methods based on quantum field theoretical treatments of the partition function \cite{inPrep}. These methods also give the correct $-1/2$ term, but in some its appearance seems simpler than others. We have chosen to give a sketch of the calculation using the method of \cite{PhysRevA.84.053633} in appendix \ref{appendixLeyronas}, since the subtleties there appear to be closely related to the ones in the spectral density method investigated here. A more comprehensive and detailed review of these issues will appear elsewhere \cite{inPrep}.
	
	\noindent This article is organized as follows: in Sec. \ref{sec:BUf} we show how a naive application of the BU formula gives an incorrect result for the 1D delta-function interaction. In Sec. \ref{sec:Spectral} we use the spectral density method to find the correction factor to the BU formula. Finally, we discuss the relation between Levinson's theorem and the spectral density method in Sec. \ref{sec:Levinson} followed by a brief discussion of the whole topic in Sec. \ref{sec:Conclusion}.   
	
	\section{Beth-Uhlenbeck formula\label{sec:BUf}}
	The BU formula \cite{beth1937quantum} connects the shift of the second virial coefficient from the free case $(\delta b_2)$ to the scattering phase shift of the potential. For a phase shift $\delta_l(k)$ corresponding to the $l$-th partial wave, the BU formula in 1D is given by [see appendix \ref{appendixbu1d}],
	\begin{equation}
	\delta b_2 = \sum_{B} e^{\beta E_B} + \frac{1}{\pi}\sum_{l=0,1} \int_{0}^{\infty}dk \hskip 0.2em e^{-\beta k^2} \frac{d\delta_l(k)}{dk},\label{BUrelation}
	\end{equation}
	where $E=-E_B$ $(E_B>0)$ is a bound state energy and $\beta = 1/k_BT$. For the 1D delta-function potential, there is only $s$-wave scattering phase shift, and hence the only possible value of $l$ is 0. The scattering phase shift for this potential is given by $\delta_0(k) = \arctan\left(\frac{\sqrt{E_B}}{k}\right)$ in the units of reduced mass $\mu= 1/2$ \cite{lapidus1969phase}. From here on we will omit the $l$ subscript on the phase shift and just denote it by $\delta(k)$. Using the BU formula given in Eq. (\ref{BUrelation}), and plugging in the phase shift of the delta-function one obtains
	\begin{equation}
	\delta b_2 = e^{\beta E_B} - \frac{1}{2} e^{\beta E_B}\left(1-\text{erf}\left[\sqrt{\beta E_B} \right]\right). 
	\end{equation} 
	It is easy to see that this expression is incorrect by checking the limit $E_B \rightarrow 0$. When the potential strength goes to zero, one should expect that the shift in the second virial coefficient vanishes. In this case, instead of obtaining zero, we get $\delta b_2 = 1/2$ in the $E_B\rightarrow 0$ limit. Clearly, this is incorrect and eq. (\ref{BUrelation}) is missing terms. To determine the missing terms, we evaluate the quantity $\delta b_2$ using the spectral density method.
	
	\section{Spectral density method \label{sec:Spectral}}
	Green's functions have extensive use in quantum mechanics \cite{economou1983green}. One of them is the usage of the spectral representation of the retarded Green's function to find out the spectral function. The spectral function denotes the probability that a certain state with momentum $k$ has energy $E$. This is exactly what the local density of states is. The local density of states can then be integrated with respected to real space to give the global density of states, a quantity which is related to scattering phase shifts. In this section, we are going to exploit this relation between the density of states and the retarded Green's function to find the missing term in the BU formula.
	
	\noindent The global density of states in the spectral density method is given by \cite{gasparian1996partial}
	\begin{equation}
	\frac{dN}{dE} = -\frac{1}{\pi}\int_{-\infty}^{\infty} dx \text{ Im}[G(x,x)], \label{dos_spectral}
	\end{equation}
	where $G(x,x')$ is the retarded Green's function of the system. For an attractive 1D delta-function potential we have exactly one bound state and a continuum of scattering states. So, the Hilbert space can be separated into two parts. One part includes the bound states and the other includes the scattering states. With this structure, the Green's function can be written as
	\begin{equation}
	G(x,x') = \lim_{\epsilon\rightarrow 0} \frac{\pb{x}\psi_B^*(x')}{E+E_B+i\epsilon} + \int_{-\infty}^{\infty} \frac{dk}{2\pi} \frac{\pk{x}\psi_k^*(x')}{E-k^2+i\epsilon},\label{gf_spectral}
	\end{equation}
	where the first term correspond to the bound state $(\pb{x})$ and the integral is over the momentum of scattering states $(\pk{x})$. In Eq. (\ref{gf_spectral}) it is implicitly understood that the scattering states are defined only for positive energy of the system ($E>0$) and bound states are defined for negative energy of the system ($E<0$).
	
	\noindent The density of states changes when the interaction is turned on. Without the interaction potential, we will only have a free particle with a continuous energy spectrum. The appearance of the bound states when the interaction is nonzero indicates the change in density of states. This change can be written in terms of Eq. (\ref{dos_spectral}),
	\begin{equation}
	\frac{d\Delta N}{dE} = -\frac{1}{\pi}\int_{-\infty}^{\infty} dx \text{ Im}[G(x,x)] + \frac{1}{\pi}\int_{-\infty}^{\infty} dx \text{ Im}[G_0(x,x)]. 
	\end{equation}
	$G_0(x,x')$ is the Green's function for the non-interacting case (free particle). The change in the density of states is related to the difference between the classical and quantum calculation of the second virial coefficient $\delta b_2$ by (see appendix \ref{appendixbu1d})
	\begin{equation}
	\delta b_2 = \int_{-\infty}^{\infty} dE \hskip 0.2em e^{-\beta E} \frac{d\Delta N}{dE}. \label{spectraldeltab2}
	\end{equation} 
	Writing out the whole expression explicitly we get
	\begin{align}
	\delta b_2 &= \int_{-\infty}^{\infty} dE \hskip 0.2em e^{-\beta E}\left(-\frac{1}{\pi}\right)\int_{-\infty}^{\infty} dx \text{ Im}[G(x,x)-G_0(x,x)]\\
	&= \int_{-\infty}^{\infty} dE \hskip 0.2em e^{-\beta E}\left(-\frac{1}{\pi}\right)\int_{-\infty}^{\infty} dx \text{ Im}\left[ \frac{|\pb{x}|^2}{E+E_B+i\epsilon} + \int_{-\infty}^{\infty} \frac{dk}{2\pi} \frac{|\pk{x}|^2- |\psi_0(x)|^2}{E-k^2+i\epsilon}\right].\label{deltab2}
	\end{align}
	where $\psi_0(x)$ is the free particle wave function (it is still dependent on $k$, but we use $\psi_0(x)$ for notational convenience). In appendix \ref{appendixSpectraltoBU} we verify that this form of writing the second virial coefficient using the spectral density function is equivalent to the BU formula. So, it may seem that this method of calculating the shift in the second virial coefficient will yield the same incorrect result for $\delta b_2$ in the case of a delta-function potential. Indeed, with a naive straightforward calculation one would end up with the same result. However, a careful consideration of divergent quantities (like the normalization of scattering states), which can explicitly be recognized in this method, yields the correct expression for $\delta b_2$ and provides a modification of the BU formula.

	\noindent To arrive at the BU formula involving phase shift from the expression with density of states, we have used a certain approximation (see appendix \ref{appendixbu1d}). First, we evaluated the number of states in a large box of volume $L$ and then took the limit $L\rightarrow \infty$. The error in following this argument is in general negligible but in certain cases this boundary condition can contribute significantly. We will see in the following discussion how this approximation contributes to the evaluation of $\delta b_2$ in the spectral density formalism.
	
	\noindent From Eq. (\ref{deltab2}), the calculation of the bound state energy part is rather straightforward and gives a contribution of $e^{\beta E_B}$ (see appendix \ref{appendixSpectraltoBU}). So, here we consider the scattering part of $\delta b_2$ and denote it by $\delta b_2^{sc}$ which is given by
	\begin{equation}
	\delta b_2^{sc} = \int_{0}^{\infty} dE \hskip 0.2em e^{-\beta E}\left(-\frac{1}{\pi}\right)\int_{-\infty}^{\infty} dx \text{ Im}\left[ \int_{-\infty}^{\infty} \frac{dk}{2\pi} \frac{|\pk{x}|^2- |\psi_0(x)|^2}{E-k^2+i\epsilon}\right].\\
	\end{equation} 
    The imaginary part of the denominator in the limit $\epsilon \rightarrow 0$ gives a delta-function in $k$ (Sokhotski-Plemelj theorem), yielding
	\begin{align}
	\delta b_2^{sc} = \int_{0}^{\infty} dE \hskip 0.2em e^{-\beta E}\int_{-\infty}^{\infty} dx \int_{-\infty}^{\infty} \frac{dk}{2\pi} \left(\left|\pk{x}\right|^2- \left|\psi_0(x)\right|^2\right) \delta(E-k^2).
	\end{align}
	We can write the scattering states for a delta-function potential as \cite{griffiths2018introduction}
	\begin{equation}
	\pk{x} = e^{ikx} + R(|k|) e^{i|k||x|}, \hskip 1em R(|k|) = -\frac{\kappa}{\kappa + i|k|},
	\end{equation}
	where $\kappa = \sqrt{E_B}$ is the wave vector corresponding to the bound state energy. After performing the integration over $k$ (using $\delta(f(x)) = \sum_{x_0} \frac{\delta(x-x_0)}{|f'(x_0)|}$, where $f(x_0)=0$) we get
	\begin{align}
	\delta b_2^{sc} = -\int_{0}^{\infty} dE \hskip 0.2em e^{-\beta E}\int_{-\infty}^{\infty} dx \frac{\sqrt{E_B}}{\pi\sqrt{E}(E+E_B)}\bigg[&\sqrt{E_B} \cos\left(\sqrt{E}(|x|-x)\right)\nonumber\\
	&+\sqrt{E}\sin\left(\sqrt{E}(|x|-x)\right)-\frac{\sqrt{
			E_B}}{2}\bigg].
	\end{align}
	The $x$ integral of the oscillating terms are divergent. The physical origin of this divergence is the non-normalizability of the scattering states. To get our way around this divergence, we introduce an infrared cutoff regularization procedure. In this process we change the limit of integration $\int_{-\infty}^{\infty} dx \rightarrow \int_{-a}^{a}dx$ with $a\rightarrow\infty$. This is conceptually the same as considering a box of length $2a$ and then taking the boundaries to infinity. We only take the limit $a\rightarrow\infty$ after performing the $E$ integral and hope that the divergences disappear. The quantity with the cutoff regularization becomes,
	\begin{align}
	\delta b_2^{sc} = -\lim_{a\rightarrow \infty}\int_{0}^{\infty} dE \hskip 0.2em e^{-\beta E}\int_{-a}^{a} dx \frac{\sqrt{E_B}}{\pi\sqrt{E}(E+E_B)}\bigg[&\sqrt{E_B} \cos\left(\sqrt{E}(|x|-x)\right)\nonumber\\
	&+\sqrt{E}\sin\left(\sqrt{E}(|x|-x)\right)-\frac{\sqrt{
			E_B}}{2}\bigg].
	\end{align}
	The integration over $x$ is now finite and can be carried out to yield
	\begin{align}
	-\lim_{a\rightarrow \infty}\int_{0}^{\infty} dE \hskip 0.2em e^{-\beta E} \frac{\sqrt{E_B}}{2\pi\sqrt{E}(E+E_B)}\left[1-\cos(2\sqrt{E}a)+\frac{\sqrt{E_B}}{\sqrt{E}}\sin(2\sqrt{E}a)\right].
	\end{align}
	The first term in the bracket is independent of the cutoff parameter $a$, and produces the same result as the BU formula, 
	\begin{equation}
	-\int_{0}^{\infty} dE \hskip 0.2em e^{-\beta E} \frac{\sqrt{E_B}}{2\pi\sqrt{E}(E+E_B)} = -\frac{
		1}{2} e^{\beta E_B}\left(1-\text{erf}\left[\sqrt{\beta E_B}\right]\right).
	\end{equation}
	So, the correction term to the BU formula appears from the cutoff regularization and needs to be evaluated explicitly. With the substitution of $E=k^2$, the term involving the cutoff parameter becomes
	\begin{align}
	 \lim_{a\rightarrow \infty}\int_{0}^{\infty} dk \hskip 0.2em e^{-\beta k^2} \frac{\sqrt{E_B}}{\pi(k^2+E_B)}\left[\cos(2ka)-\frac{\sqrt{E_B}}{k}\sin(2ka)\right]. \label{cutoffcorrection}
	\end{align}
	The first term within the brackets can be rewritten as
	\begin{equation}
	\lim_{a\rightarrow \infty}\frac{1}{2\pi} \int_{-\infty}^{\infty} dk \hskip 0.2eme^{-\beta k^2} \frac{\sqrt{E_B}}{(k^2+E_B)} e^{2ika} = 0.\label{cutoffcos}
	\end{equation}
	The integration can be done by considering a semicircular contour in the upper half plane because $a>0$ (Fig. \ref{fig:contours}(a)). The pole $k=+i\sqrt{E_B}$ contributes to the residue and in the limit $a\rightarrow \infty$, the residue vanishes. The second term in Eq. (\ref{cutoffcorrection}), however, has three poles, at $k=\pm i \sqrt{E_B}$ and $k=0$. We consider a similar semicircular contour in the upper half plane but now there is another pole on the real axis $k=0$ (Fig. \ref{fig:contours}(b)). To avoid this, we take an infinitesimal semicircular detour around it (with radius $\epsilon\rightarrow 0$). The contribution from the pole inside the contour gives a zero contribution in the limit $a\rightarrow\infty$ like before but the small infinitesimal arc contributes a $-\frac{1}{2}$ term,
	\begin{equation}
	-\lim_{a\rightarrow \infty}\frac{1}{2\pi i} \int_{-\infty}^{\infty} dk \hskip 0.2eme^{-\beta k^2} \frac{E_B}{k(k^2+E_B)} e^{2ika} = -\frac{1}{2}. \label{cutoffsin}
	\end{equation}
	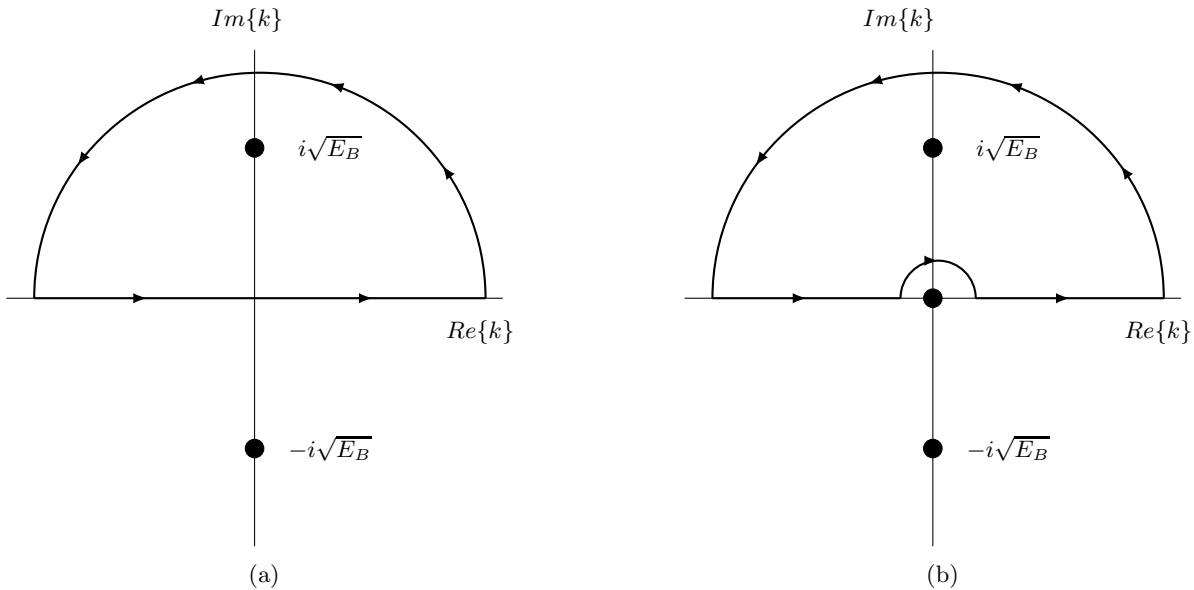
\begin{figure}
\centering
\begin{subfigure}{0.47\textwidth}
\begin{tikzpicture}
\def\gap{0}
\def\bigradius{2.6}
\def\littleradius{0.5}

\draw (-3.3, 0) -- (3.3,0)
      (0, -3.3) -- (0, 3.3);
\draw[thick,black,xshift=2pt,
decoration={ markings,  
      mark=at position 0.25 with {\arrow{latex}}, 
      mark=at position 0.75 with {\arrow{latex}}}, 
      postaction={decorate}]
  (-3,0) -- (3,0);
\draw[thick,black,xshift=2pt,
decoration={ markings,
      mark=at position 0.2 with {\arrow{latex}}, 
      mark=at position 0.4 with {\arrow{latex}},
      mark=at position 0.6 with {\arrow{latex}}, 
      mark=at position 0.8 with {\arrow{latex}}}, 
      postaction={decorate}]
 (3,0) arc (0:180:3) -- (-3,0);
 
\begin{scriptsize}
\draw [fill=black] (0,2)circle (3.5pt);
\draw [fill=black] (0,-2)circle (3.5pt);
\end{scriptsize}
\node at (3,-0.45){$Re\{k\}$};
\node at (-0.1,3.7) {$Im\{k\}$};

\node at (1,2) {$i\sqrt{E_B}$};
\node at (1,-2) {$-i\sqrt{E_B}$};
\end{tikzpicture}
\caption{} \label{ta}
\end{subfigure}
\hspace*{\fill}
\begin{subfigure}{0.47\textwidth}
\begin{tikzpicture}
\def\gap{0}
\def\bigradius{2.6}
\def\littleradius{0.5}

\draw (-3.3, 0) -- (3.3,0)
      (0, -3.3) -- (0, 3.3);
\draw[thick,black,xshift=2pt,
decoration={ markings,  
      mark=at position 0.5 with {\arrow{latex}}}, 
      postaction={decorate}]
  (-3,0) -- (-0.5,0);
  \draw[thick,black,xshift=2pt,
decoration={ markings,  
      mark=at position 0.5 with {\arrow{latex}}}, 
      postaction={decorate}]
  (0.5,0) -- (3,0);
\draw[thick,black,xshift=2pt,
decoration={ markings,
      mark=at position 0.2 with {\arrow{latex}}, 
      mark=at position 0.4 with {\arrow{latex}},
      mark=at position 0.6 with {\arrow{latex}}, 
      mark=at position 0.8 with {\arrow{latex}}}, 
      postaction={decorate}]
 (3,0) arc (0:180:3) -- (-3,0);
 
 \draw[thick,black,xshift=2pt,
decoration={ markings,
      mark=at position 0.5 with {\arrow{latex}}}, 
      postaction={decorate}]
 (-0.5,0) arc (180:0:0.5) -- (0.5,0);
 
\begin{scriptsize}
\draw [fill=black] (0,0) circle (3.5pt);
\draw [fill=black] (0,2)circle (3.5pt);
\draw [fill=black] (0,-2)circle (3.5pt);
\end{scriptsize}
\node at (3,-0.45){$Re\{k\}$};
\node at (-0.45,3.7) {$Im\{k\}$};

\node at (1,2) {$i\sqrt{E_B}$};
\node at (1,-2) {$-i\sqrt{E_B}$};
\end{tikzpicture}
\caption{} \label{tb}
\end{subfigure}
\hspace*{\fill}
\caption{(a) Contour for doing integration in Eq. (\ref{cutoffcos}). The poles are at $k=\pm i\sqrt{E_B}$ but there is no pole at $k=0$. (b)  Integration contour required for in Eq. (\ref{cutoffsin}). The poles are at $k=\pm i\sqrt{E_B}$ and at $k=0$. The small detour around $k=0$ contributes the extra $-1/2$ term.}
		\label{fig:contours}
	\end{figure}
	
	\noindent Combining Eq. (\ref{cutoffcos}), (\ref{cutoffsin}) and the contribution from the bound state part we now have the complete result,
	\begin{equation}
	\delta b_2 = e^{\beta E_B} -\frac{
		1}{2} e^{\beta E_B}\left(1-\text{erf}\left[\sqrt{\beta E_B}\right]\right) - \frac{1}{2} . \label{deltab2corrected}
	\end{equation}
	We see that in the limit $E_B\rightarrow 0$, this now gives the correct answer which is $\delta b_2 = 0$. The $-1/2$ term came from the residue of the pole at $k=0$, which is equivalent to the contribution of the zero-energy state. It shows that we must be careful while dealing with the boundary terms that produces a divergence (taking the limit that the box length goes to infinity). 

	\section{Correction to BU formula in 1D and Levinson's theorem \label{sec:Levinson}}
	In the previous section we found that there are subtle correction terms to the BU formula appearing from the infrared regularization procedure and carefully taking the length of the box to infinity. So, where did it go wrong in the BU formula derivation? As shown in appendix \ref{appendixSpectraltoBU}, the scattering part of Eq. (\ref{deltab2}) can also be rewritten as
	\begin{equation}
	\delta b_2^{sc} = \int_{0}^{\infty} \frac{dk}{\pi} \hskip 0.2em e^{-\beta k^2}  \int_{-\infty}^{\infty} dx \left(|\pk{x}|^2- |\psi_0(x)|^2\right)\label{probdensitymain}.
	\end{equation}
	Looking closely at Eq. (\ref{probdensitymain}), one can recognize that the $x$ integral over the scattering states are divergent as well. Instead of treating this divergence carefully, we rewrote the probability density (density of states) in terms of phase shifts  assuming that there is no consequence of ignoring the effects of taking the box boundaries to infinity. In fact, one can carry out the same cutoff regularization scheme starting from Eq. (\ref{probdensitymain}) (considering $\pk{x} = \cos(k|x|+\delta(k))$) and get a more generalized correction term involving phase shifts equivalent to Eq. (\ref{cutoffcorrection}). This approach gives us the modification to the BU formula
	\begin{gather}
	\delta b_2 = \sum_{B} e^{\beta E_B} + \int_{0}^{\infty} \frac{dk}{\pi} \frac{d\delta(k)}{dk}\hskip 0.2em e^{-\beta k^2} + I,\label{BUcorrected}\\
	\text{where } I = \lim_{a\rightarrow \infty}\int_{0}^{\infty} dk \hskip 0.2em \frac{e^{-\beta k^2}}{2\pi k}\left[\sin(2ka)(\cos(2\delta(k))-1) + \cos(2ka)\sin(2\delta(k))\right] \label{correctionphaseshift}
	\end{gather}
	One can check that Eq. (\ref{correctionphaseshift}) and Eq. (\ref{cutoffcorrection}) are identical by plugging in the phase shift for the delta-function potential $\delta(k) = \arctan\left(\frac{\sqrt{E_B}}{k}\right)$. The contribution of the correction term depends on the nature of the interacting system at $k=0$. For a system with a bound state at zero energy ($k=0$), the integral $I$ vanishes in the limit $a\rightarrow \infty$ (because there is no pole anymore at $k=0$). Otherwise it will contribute a nonzero correction to Eq. (\ref{BUcorrected}). This statement has a very close relation to Levinson's theorem \cite{levinson1949determination,levinson1949uniqueness,newton1977noncentral,wellner1964levinson}. Levinson's theorem relates the number of bound states for a symmetric potential to the phase shift at zero energy and infinite energy. 
	
	\noindent To see how the spectral density formalism relates to Levinson's theorem in 1D, we refer back to Eq. (\ref{dos_spectral}). Instead of finding $\delta b_2$, now we want to calculate $\Delta N$ by integrating over all energy. Integrating over all energies should give $\Delta N = 0$ because the introduction of the potential only changes the density of states, but we still are within the same Hilbert space \cite{weinberg2015lectures}. From Eq. (\ref{dos_spectral}), integrating both sides w.r.t. $E$, we obtain
	\begin{equation}
	\Delta N = \int_{-\infty}^{\infty} dE \hskip 0.2em \left(-\frac{1}{\pi}\right)\int_{-\infty}^{\infty} dx \text{ Im}\left[\sum_B \frac{|\pb{x}|^2}{E+E_B+i\epsilon} + \int_{-\infty}^{\infty} \frac{dk}{2\pi} \frac{|\pk{x}|^2- |\psi_0(x)|^2}{E-k^2+i\epsilon}\right] \label{deltaN}.
	\end{equation}
	Here, we have introduced a sum over the number of bound states in the expression of the Green's function. In case of the attractive delta-function potential there is just one bound state, but in general there can be more than one bound state for a given potential. In Eq. (\ref{deltaN}) the $k$ and $x$ integration can be carried out in the same way as in Sec. \ref{sec:Spectral} to arrive at the result
	\begin{equation}
	\Delta N = \sum_{B}\int_{-\infty}^{0}dE \hskip 0.2em \delta(E+E_B) + \frac{1}{\pi}\int_{0}^{\infty} dk \hskip 0.2em \frac{d\delta}{dk} + I_1,
	\end{equation}
	where
	\begin{equation}
	I_1 = \lim_{a\rightarrow \infty}\int_{0}^{\infty} dk \hskip 0.2em \frac{1}{2\pi k}\left[\sin(2ka)(\cos(2\delta(k))-1) + \cos(2ka)\sin(2\delta(k))\right].
	\end{equation} 
	Carrying out the E integral on the RHS now gives
	\begin{align}
	\Delta N &= N_B + \frac{\delta(\infty)-\delta(0)}{\pi} + I_1\\
	\implies N_B &= \frac{\delta(0)-\delta(\infty)}{\pi} - I_1. \label{Levinson}
	\end{align}
	where $N_B$ is the number of bound states. Eq. (\ref{Levinson}) is precisely the Levinson's theorem in 1D for the even parity case which relates the number of bound states to the scattering phase shift. For a potential with $E_B> 0$ ($E=-E_B<0$), $I_1$ is non-zero, otherwise for a half bound state ($E_B=0$) it is zero. In the case of delta-function potential which does not support any half bound state, $I_1 = -1/2$, and $\delta(0) = \pi/2$, giving $N_B = 1$, which is exactly the number of bound states supported by the potential.
	
	\noindent One can show Levinson's theorem in 1D by using a number of methods, including Sturm-Liouville method, and the $S$-matrix method \cite{dong2000levinson,sassoli1994levinson,barton1985levinson,eberly1965quantum}. From those calculations one ends up with the following form of Levinson's theorem (in the even parity case),
	\begin{equation}
	N_B = \begin{cases}
	\frac{\delta(0)-\delta(\infty)}{\pi} \text{ \hskip 2.9em for critical case}\\
	\frac{\delta(0)-\delta(\infty)}{\pi} + \frac{1}{2} \text{ \hskip 1em for non-critical case}.
	\end{cases}
	\end{equation}
	where by the critical case we mean the case with $E_B = 0$ (half bound state) and non-critical case means for $E_B>0$. So, we see that the value of the correction term $I_1$ is actually universal and is equal to $-1/2$ (for $E_B>0$) irrespective of the exact form of the potential. We have used the even parity $(\pk{x} = \cos(k|x|+\delta(k)))$ case to arrive at Levinson's theorem and the correction term to the BU formula using the spectral density method. One can similarly use an odd parity scattering wave function and follow the same regularization procedure to arrive at the odd parity results of the Levinson's theorem as well. 
	
	\section{Discussion \label{sec:Conclusion}}
	Cutoff regularization schemes, both ultraviolet (high energy/momentum) and infrared (large volume/small energy/momentum) are highly used in quantum field theory to deal with infinities in the calculations. In this note we saw how using an infrared cutoff can yield the correction term in the original BU formula in 1D,  and leads to the unusual form of Levinson's theorem in this case. Compared to other methods, the spectral density method provides a rather straightforward and physically insightful way to derive the extra $-1/2$ correction  term in the BU formula, as well as the corresponding Levinson's theorem in 1D.  We also use a quantum field theory method that shows that it is only the zero-energy behavior of the theory, regardless of the actual functional form of the potential, that dictates these correction terms; and hence we see an apparent surprising universality of the values of the integrals $I$ and $I_1$. 

	\section*{Acknowledgment}
	This work was supported in part by the U.S. National Science Foundation  under  Grant  No. PHY1452635 (Computational Physics Program), the US Army Research Office Grant No. W911NF-15-1-0445, and the University of San Francisco Faculty Development Fund.
	
	\appendix
	
	\section{Virial Expansion with Feynman Diagrams\label{appendixLeyronas}}{

\setlength\parindent{0pt}

\newcommand{\da}{\dagger}
\newcommand{\f}[2]{\frac{#1}{#2}}
\newcommand{\ra}{\rangle}
\newcommand{\la}{\langle}
\newcommand{\p}{\partial}
\newcommand{\Q}{\left}
\newcommand{\W}{\right}
\newcommand{\en}{\begin{equation}\begin{aligned}}
\newcommand{\een}{\end{aligned} \end{equation}}

Various techniques have been developed for perturbatively calculating virial coefficients from field theory \cite{PhysRev.187.345,PhysRevA.84.053633,PhysRevLett.107.030601,PhysRevA.91.013606}. In particular, we will employ the technique developed in \cite{PhysRevA.84.053633}, which expands the many-body propagator in powers of fugacity and works directly in imaginary time instead of frequency. One can also use the field theoretic method described in \cite{daza2018virial,chafin2013scale} where the Hubbard-Stratonovich transformation has been used to get the $-1/2$ term. However, this following method includes some subtleties that resemble the spectral density method discussed in the main paper.\\

The noninteracting many-body time-ordered propagator is given by the well-known expression \cite{abrikosov1975methods}

\en
G^0(p,\tau)&= e^{-(\epsilon_p-\mu) \tau}\Q(-\Theta(\tau)+n_F(\epsilon_p-\mu)\W)\\
&=e^{\mu \tau} \Q[\sum_{n \geq 0} G^{(0,n)}(p,\tau)z^n\W],
\een

where we have made a virial expansion of the Fermi distribution function $n_F(x)=(e^x+1)^{-1}$ to define:

\en
G^{(0,0)}(p,\tau)&=-\Theta(\tau)e^{-\epsilon_p \tau}\\
G^{(0,n\geq1)}(p,\tau)&=(-1)^{n-1}e^{-\epsilon_p \tau}e^{- n \beta \epsilon_p}.
\een

The basic procedure is to note that $n=\f{\p P}{\p \mu}\big |_\beta=-\lim \limits_{\eta \rightarrow 0-}\int \f{dp}{2\pi} \,G(p,\eta)$, since

\en
 G(0,\eta)&= \la\psi (x,\eta)\psi^\da (x,0 )\ra\\
 \lim \limits_{\eta \rightarrow 0-} G(0,\eta)&=-\la\psi^\da (x,0 )\psi (x,0)\ra=-n\\
 &=\lim \limits_{\eta \rightarrow 0-}\int \f{dp}{2\pi} \,G(p,\eta).
\een

Therefore knowledge of the exact 2-pt Green's function $G(p,\tau)$ gives the number density from which we can extract $\delta b_2$. The exact 2-pt Green's function is built perturbatively from the noninteracting 2-pt Green's function, so a controlled expansion in fugacity can be obtained. The contribution to  $\delta b_2$ can be readily seen to be given by just a single diagram given in Fig \ref{chrisFig1},
\begin{figure}[h] 
\centering
\begin{tikzpicture}

\draw[->,thick] (0,0) -- (4,0) coordinate (xaxis)
node[pos=0.25]{$|$} node[pos=0.5]{$|$}node[pos=0.75]{$|$}
node[pos=0.375,below]{$t_1$} node[pos=0.625,below]{$t_2$};

\node[right] at (xaxis) {$\tau$};

\draw[->,thick] (1,0)--(1.5,.75);
\draw[thick] (1.4,.60)--(2,1.5);

\node at (1.15,1) {${k \! \downarrow}$};

\draw[->,thick] (1,0)--(2,.75);
\draw[thick] (1.9,.675)--(3,1.5)
node[pos=.55]{$\mathbf{ \bracevert}$};

\node at (3,1) {${k  \! \downarrow}$};

\draw[ultra thick] (2,1.5) rectangle (3,2.5);
\filldraw[color=gray] (2,1.5) rectangle (3,2.5);

\node at (2.5,2){${T_2}$};

\draw[thick,->] (3,2.5)..controls (4.5,3.7)  and (.5,3.7) ..(2,2.5)
node[pos=.5]{$\mathbf{ \bracevert}$};

\node at (2.5,3){${P-k  \! \uparrow}$};

\end{tikzpicture}

\caption{Diagram corresponding to the calculation of $\delta b_2$. The tree-level contact coupling is replaced by the full 1PI 2-body T-matrix $T_2$, where $T_2$ is a function only of the center of mass momentum $P$ flowing through the s-channel. The tick marks represent $G^{(0,1)}$ propagators.}\label{chrisFig1} 

\end{figure}
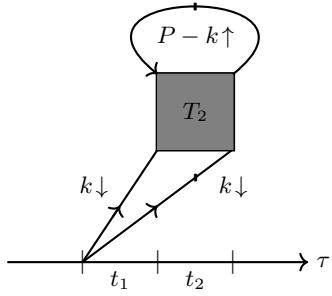
corresponding to the expression

\en
\delta n_2=-\int \f{dP}{2\pi} \f{dk}{2\pi}dt_1dt_2 \,\theta(t_1)\theta(t_2)\theta(\beta-t_1-t_2)T_2(P,t_2)e^{-(\beta-t_2)(\epsilon_k+\epsilon_{P-k})},
\een

where the only nonvanishing integration region is $t_1,t_2>0$ with  $t_1+t_2<\beta$, $T_2$ is the off-shell 2-body T-matrix analytically continued to imaginary time, and $\theta(t)$ is the Heaviside step function. Integrating over $t_1$ gives

\en \label{intermediateEqn}
\delta n_2&=-\int \f{dP}{2\pi} \f{dk}{2\pi}dt_2 \, \theta(t_2) T_2(P,t_2)\theta(\beta-t_2) (\beta-t_2) e^{-(\beta-t_2)(\epsilon_k+\epsilon_{P-k})}\\
&=-\int \f{dP}{2\pi} \f{dk}{2\pi}\f{ds}{2\pi i}e^{-\beta s}T_2(s-P^2/4)\f{1}{(s-\epsilon_k-\epsilon_{P-k})^2},
\een

where we used that the inverse Laplace transform of a product is a convolution $\int dt \,g(t)f(\beta-t)=\int \f{1}{2\pi i} e^{-\beta s} f(s)g(s)$, $\f{1}{(s-\epsilon_k-\epsilon_{P-k})^2}$ is the Laplace transform of $t e^{-t(\epsilon_k+\epsilon_{P-k})}$, $T_2(s-P^2/4)=\f{C}{1+\f{C}{2}\f{1}{\sqrt{-(s-P^2/4)-i\epsilon}}}$ is the Euclidean 2-body T-matrix in the usual frequency space, and $C$ is the coupling constant. It should be mentioned that the usual Bromwich-Mellin contour over which the integral over $s$ is taken begins at $\text{Re}[s]<0$ as the sign of the exponential in the Laplace transform is taken opposite the standard convention in mathematics \cite{arfken2013mathematical}. A consequence of this is that the integrand of \eqref{intermediateEqn} is analytic to the left of the contour instead of the usual right of the contour, and in evaluating the integral we close the contour on the right instead of the left, as indicated in Fig \ref{chrisFig2}.\\

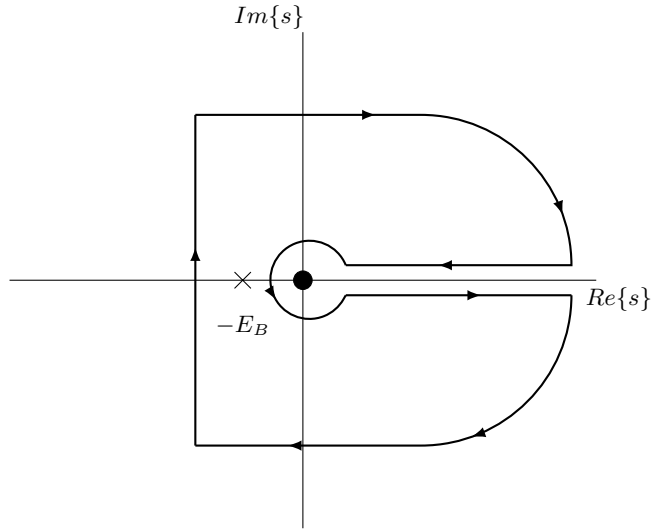
\begin{figure}[h] 
\centering
\begin{tikzpicture}
\def\gap{0.2}
\def\bigradius{3}
\def\littleradius{0.5}

\draw (-1.3*\bigradius, 0) -- (1.3*\bigradius,0)
      (0, -1.1*\bigradius) -- (0, 1.1*\bigradius);
\draw[thick,black,xshift=2pt,
decoration={ markings,  
      mark=at position 0.8 with {\arrow{latex}}}, 
      postaction={decorate}]
  (-1.5,2.2) -- (1.5,2.2);
  
\draw[thick,black,xshift=2pt,
decoration={ markings,
      mark=at position 0.4 with {\arrow{latex}},
      mark=at position 0.8 with {\arrow{latex}}}, 
      postaction={decorate}]
 (1.5,2.2) arc (90:0:2) -- (0.5,0.2);
 \draw[thick,black,xshift=2pt,
decoration={ markings,
      mark=at position 0.4 with {\arrow{latex}},
      mark=at position 0.8 with {\arrow{latex}}}, 
      postaction={decorate}]
 (3.5,-0.2) arc (0:-90:2) -- (-1.5,-2.2);

   \draw[thick,black,xshift=2pt,
decoration={ markings,  
      mark=at position 0.6 with {\arrow{latex}}}, 
      postaction={decorate}]
  (0.5,0.2) arc (22:330:0.52)-- (0.5,-0.2);
    \draw[thick,black,xshift=2pt,
decoration={ markings,  
      mark=at position 0.6 with {\arrow{latex}}}, 
      postaction={decorate}]
  (0.5,-0.2) -- (3.5,-0.2);

   \draw[thick,black,xshift=2pt,
decoration={ markings,  
     mark=at position 0.6 with {\arrow{latex}}}, 
      postaction={decorate}]
  (-1.5,-2.2) -- (-1.5,2.2);

\begin{scriptsize}
\draw [fill=black] (0,0) circle (3.5pt);
\draw (-0.8,0) node[cross] {};
\end{scriptsize}
\node at (4.2,-0.25){$Re\{s\}$};
\node at (-0.45,3.5) {$Im\{s\}$};

\node at (-0.8,-0.6) {$-E_B$};
\end{tikzpicture}

\caption{Contour for evaluating the Bromwich integral. Besides the pole at the bound state $s=-E_B$ and the branch cut for $s>0$, a singularity exists at the origin that goes as $1/s$, so that the infinitesimal circle surrounding the origin contributes to the integral.}\label{chrisFig2} 
\end{figure}

 Using the substitution $k\rightarrow k+P/2$ followed by $s \rightarrow s+P^2/4$, the integrals over $k$ and $P$ become trivial and we get:

\en
\delta n_2&=\f{i}{8\pi^{3/2}\sqrt{\beta}} \int ds\,e^{-\beta s}\f{1}{(-s)^{3/2}}
\f{C}{1+\f{C}{2}\f{1}{\sqrt{-s}}}.
\een

The analytic structure is such that $\f{1}{1+\f{C}{2}\f{1}{\sqrt{-s}}}$ has a pole at the bound state $s=-C^2/4=-E_b$ for $C<0$ with residue $-\f{C^2}{2}$, and $\f{1}{(-s)^{3/2}}$ has a branch cut along the positive real axis $s>0$. Therefore closing the contour on the right half of the complex $s$-plane, the pole gives a contribution

\en \label{eqnBoundN}
\delta n^{\text{bound}}_2&=(-2\pi i)\f{i}{8\pi^{3/2}\sqrt{\beta}}\f{1}{\Q(\f{|C|}{2}\W)^3} e^{\beta C^2/4}C \Q(-\f{C^2}{2}\W)\\
&=\f{1}{\pi^{1/2}\sqrt{\beta}}e^{\beta E_b},
\een

while the branch cut gives the contribution

\en \label{eqnScatN}
\delta n^{\text{scattering}}_2&=\f{i}{8\pi^{3/2}\sqrt{\beta}} \int_0^\infty dx\, e^{-\beta x} \text{disc}\Q[\f{1}{(-x-i\epsilon)^{3/2}}
\f{C}{1+\f{C}{2}\f{1}{\sqrt{-s-i\epsilon}}}
\W]\\
&=\f{i}{8\pi^{3/2}\sqrt{\beta}}  \int_0^\infty dx\, e^{-\beta x} (2 i )\text{ Im}\Q[\f{1}{(-x-i\epsilon)^{3/2}}
\f{C}{1+\f{C}{2}\f{1}{\sqrt{-s-i\epsilon}}}
\W]\\
&=\f{i}{8\pi^{3/2}\sqrt{\beta}}  \int_0^\infty dx\, e^{-\beta x} (2 i )\Q[-\f{1}{x^{3/2}}\f{C}{1+\f{C^2}{4x}}\W]\\
&=-\f{1}{2 \pi^{1/2}\sqrt{\beta}}e^{\beta E_b}\Q[1-\text{erf}\Q(\sqrt{\beta E_b}\W)\W].
\een

Additionally, the infinitesimal circle surrounding the origin does not vanish and contributes

\en \label{eqnoriginN}
\delta n^{\text{origin}}_2&=-\f{i}{8\pi^{3/2}\sqrt{\beta}} \lim_{R\rightarrow 0}\int d(Re^{i\theta}) \,e^{-\beta (Re^{i\theta})}
\f{1}{(-Re^{i\theta})^{3/2}}
\f{C}{1+\f{C}{2}\f{1}{\sqrt{-Re^{i\theta}}}}\\
&=
-\f{i}{8\pi^{3/2}\sqrt{\beta}} \lim_{R\rightarrow 0}\int_0^{2\pi} (i Re^{i\theta})d\theta \,
\f{1}{(-Re^{i\theta})^{3/2}}
\f{C}{\f{C}{2}\f{1}{\sqrt{-Re^{i\theta}}}}\\
&=-\f{1}{2\pi^{1/2}\sqrt{\beta}}.
\een

Adding Eqs. \eqref{eqnBoundN}, \eqref{eqnScatN}, and \eqref{eqnoriginN}, one can read off:

\en
\delta b_2=\f{1}{2}e^{\beta E_b}\Q(1+\text{erf}(\sqrt{\beta E_b})\W)-\f{1}{2}
\een

Note that if the circle around the origin were not included, one would be missing a $1/2$. In 2D and 3D, this circle gives a vanishing contribution. The origin of the term $-1/2$ in this method is very similar to the origin of the term spectral density method, both given by a contribution of the contour around the pole at zero energy in the complex plane.
}
\section{Second virial coefficient and density of states \label{appendixbu1d}}
{
\setlength\parindent{0pt}

\newcommand{\e}{\begin{equation*}\begin{aligned}}
\newcommand{\ee}{\end{aligned}\end{equation*}}
\newcommand{\en}{\begin{equation}\begin{aligned}}
\newcommand{\een}{\end{aligned} \end{equation}}
\newcommand{\fp}[2]{\frac{d^{#2} #1}{(2\pi)^{#2}}}
\newcommand{\pfa}[2]{\frac{\delta #1}{\delta #2}}
\newcommand{\p}{\partial}
\newcommand{\pf}[2]{\frac{\p #1}{\p #2}}
\newcommand{\f}[2]{\frac{#1}{#2}}
\newcommand{\ra}{\rangle}
\newcommand{\la}{\langle}
\newcommand{\da}{\dagger}
\newcommand{\ma}{\mathcal}
\newcommand{\tr}{\text{tr }}
\newcommand{\Q}{\left}
\newcommand{\W}{\right}
\newcommand{\pma}{\begin{pmatrix}}
\newcommand{\epma}{\end{pmatrix}}
\newcommand{\na}{\nabla}
\newcommand{\de}{\delta}
\newcommand{\ep}{\epsilon}
\newcommand{\Tr}{\text{tr}}

The asymptotic form for a scattering wavefunction in 1D is 

\en
\psi_\text{\textbf{k}}(x)=e^{i\mathbf{k} x}+i e^{ik|x|}f(\text{sgn} (x),k),
\een

which has a partial wave decomposition \cite{doi:10.1119/1.1970982}

\en
\psi_\text{\textbf{k}}(x)=\sum_{\ell=0,1}(-i)^\ell (\text{sgn}(x))^\ell e^{i \delta_\ell }\cos(k |x|+\zeta_\ell+ \delta_\ell),
\een

where $\zeta_\ell=\pi \ell/2$, $f(\text{sgn} (x),k)=\sum \limits_{\ell=0,1}(\text{sgn}(x))^\ell e^{i \delta_\ell} \sin \delta_\ell$.

Imposing hard-wall boundary conditions at $|x|=R$ quantizes $k$

\en
k_nR+\zeta_\ell+ \delta_\ell(k_n)=(n+1/2)\pi,
\een

and assuming that, for large $R$, the phase shift becomes a smooth function of $k$ then
\en
(k_{n+1}-k_n)R+ (k_{n+1}-k_n)\delta'_\ell(k_n)&=\pi\\
\f{dN_\ell}{dk}&=\f{R+\delta'_\ell(k)}{\pi}\\
\f{d\Delta N_\ell}{dk}&=\f{\delta'_\ell(k)}{\pi}.
\een

The definition of the relative 2-body partition function in terms of the density of states $g(E)$ is
\en
Z_2=\int dE \,g(E) e^{-\beta E}.
\een

Therefore
\en
\Delta Z_2=Z_2-Z^{\text{0}}_2&=\sum_{\ell=0}^1 \sum_{i=0}^{n_\ell} e^{\beta E_{b,\ell,i}}+ \sum_{\ell=0}^1 \int dE \f{d\Delta N_\ell}{dE} e^{-\beta E}\\
&=\sum_{\ell=0}^1 \sum_{i=0}^{n_\ell} e^{\beta E_{b,\ell,i}}+\f{1}{\pi} \sum_{\ell=0}^1 \int_0^\infty dk \, \f{d \delta_\ell}{dk}e^{-\beta k^2},
\een
where we used the bound state part $g_{\text{bound}}(E)=\sum \limits_{\ell=0}^1\sum  \limits_{i=0}^{n_\ell} \delta(E-E_{b,\ell,i})$ and defined $n_\ell$ to be the number of bound states of parity $\ell$.\\
In this paper we take as our convention for $\Delta b_2$:

\en
\Delta b_2=\Delta Z_2=\sum_{\ell=0}^1 \sum_{i=0}^{n_\ell} e^{\beta E_{b,\ell,i}}+\f{1}{\pi} \sum_{\ell=0}^1 \int_0^\infty dk \, \f{d \delta_\ell}{dk}e^{-\beta k^2}.
\een
}

\section{From spectral density method to BU formula \label{appendixSpectraltoBU}}
	In this appendix we verify that using the retarded Green's function to find density of states is equivalent to the BU formula. To show this, we start from equation Eq. (\ref{deltab2})
	\begin{equation}
	\delta b_2= \int_{-\infty}^{\infty} dE \hskip 0.2em e^{-\beta E}\left(-\frac{1}{\pi}\right)\int_{-\infty}^{\infty} dx \text{ Im}\left[\sum_B \frac{|\pb{x}|^2}{E+E_B+i\epsilon} + \int_{-\infty}^{\infty} \frac{dk}{2\pi} \frac{|\pk{x}|^2- |\psi_0(x)|^2}{E-k^2+i\epsilon}\right].\label{appendixdeltab2}
	\end{equation}
	Here we have introduced a sum over bound states as well since for a more general potential there can be more than one bound state. Now we will rearrange the orders of integration. Since the continuum part displays some complexity, we will come back to it after we compute the contribution of the discrete bound states part first.
	
	The integral over the bound states is
	\begin{align}
	&\int_{-\infty}^{0} dE \hskip 0.2em e^{-\beta E}\left(-\frac{1}{\pi}\right)\int_{-\infty}^{\infty} dx \text{ Im}\left[\sum_B \frac{|\pb{x}|^2 }{E+E_B+i\epsilon}\right]\\
	=&\int_{-\infty}^{0} dE \hskip 0.2em e^{-\beta E}\left(-\frac{1}{\pi}\right)(-\pi) \sum_B \delta(E+E_B)\\
	=&\sum_B e^{\beta E_B}.
	\end{align} 
	The limit of the integration is $-\infty$ to $0$ because the bound states are only defined for negative energy. To arrive at the second line we have used the normalization of the bound states $\int_{-\infty}^{\infty}|\pb{x}|^2 dx = 1$ and the delta-function appears due to Sokhotski-Plemelj theorem. The appearance of the factor $\delta(E+E_B)$ can also be visualized by the fact that the imaginary part of the factor inside the bracket is zero everywhere except only when $E=-E_B$ and at $E=-E_B$ the factor is infinite which mimics the definition of the delta-function.
	
	Coming back to the continuum part of Eq. (\ref{deltab2}), we can integrate with respect to $E$ first to get
	\begin{align}
	=& \left(-\frac{1}{\pi}\right)\int_{-\infty}^{\infty} dx \int_{-\infty}^{\infty} \frac{dk}{2\pi}\left(|\pk{x}|^2- |\psi_0(x)|^2\right) \text{ Im}\left[\int_{0}^{\infty} dE \hskip 0.2em \frac{e^{-\beta E}}{E-k^2+i\epsilon}\right]\\
	=& \int_{-\infty}^{\infty} dx \int_{-\infty}^{\infty} \frac{dk}{2\pi}\left(|\pk{x}|^2- |\psi_0(x)|^2\right) e^{-\beta k^2}\\
	=& \int_{0}^{\infty} \frac{dk}{\pi} \hskip 0.2em e^{-\beta k^2}  \int_{-\infty}^{\infty} dx \left(|\pk{x}|^2- |\psi_0(x)|^2\right). \label{probdensity}
	\end{align}
	The $x$ integral gives the change in probability density for a particular momentum $k$, which is nothing but the density of states $g(k)$. Recognizing this reduces Eq. (\ref{probdensity}) to the familiar form of the BU formula derivation (Eq. (\ref{BUdos})), which can subsequently be rewritten in terms of the scattering phase shifts, giving the celebrated BU formula.  
	\begin{align}
	\delta b_2 &= \sum_{B} e^{\beta E_B} + \int_{0}^{\infty} \frac{dk}{\pi} \hskip 0.2em e^{-\beta k^2} \left[g(k) - g_0(k)\right] \label{BUdos}\\
	&= \sum_{B} e^{\beta E_B} + \int_{0}^{\infty} \frac{dk}{\pi} \frac{d\delta(k)}{dk}\hskip 0.2em e^{-\beta k^2} \label{BUformula}
	\end{align}
	Here in writing the second line we have used the result from appendix \ref{appendixbu1d}.

\end{document}